# NI-MH BATTERY MODELLING FOR AMBIENT INTELLIGENCE APPLICATIONS

*D. Szente-Varga, Gy. Horvath, M. Rencz*

(szvdom | horvath | rencz@eet.bme.hu)
Budapest University of Technology and Economics
Department of Electron Devices

## ABSTRACT

Mobile devices, like sensor networks and MEMS actuators use mobile power supplies to ensure energy for their operation. These are mostly batteries. The lifetime of the devices depends on the power consumption and on the quality and capacitance of the battery. Though the integrated circuits and their power consumption improve continually, their clock frequency also increases with the time, and the resultant power consumption seems not to vary, or slightly increase. On the other hand, the properties of batteries are developing much slower, necessitating the optimization of their usage on system level.

## 1. INTRODUCTION

A possible solution to increase the lifetime of mobile applications is to consider battery-aware design and practice [1] [2] [3]. It is characteristic to each type of the batteries, that the total charge that could be drawn out from the battery during a whole discharge period depends drastically on the duration of the discharge time period (how fast the charge was drawn out), and also on the waveform of the discharging current. Constant current discharging and using several types of pulsing discharge currents results different lifetimes of the battery, however the total charge in each time unit is equal.

Extending the battery's lifetime is nowadays not only the task of hardware developers, who can improve technology and design and manufacture devices with even lower and lower power consumption. Today even software developers have to take also the battery-aware system designing rules into consideration. It may result in significantly different battery lifetime. If the running application software results in constant or pulsing power consumption, or perhaps rarely generates a high current impulse and in the other longer time period no current is drawn out. These different types of power consumption rates represent different hardness of loads for the battery, even if they have got the same sum of charge consumed in a longer time unit. Knowing these differences, the operating software has to be designed to assure reaching the batteries longest lifetime.

To examine these effects, to make comparisons and eventually for the simulations there is a need for a proper battery model. In this paper a new model will be presented, which takes the magnitude and time-variation of the battery's load into consideration. This model is enabling simulations with constant current and several different mode pulsing current loads.

## 2. PREVIOUSLY PREPARED MODEL

Last year we have presented a battery model that was aimed at modeling the Ni-MH battery's behavior depending on the environmental temperature and on the load [4]. The batteries have been measured under constant resistant loads and at constant temperature values, which means, during each discharge cycle the load did not vary. No pulsing or impulse mode discharging behavior was examined.

The batteries were measured at different temperature values, namely 1°C, 5°C, 10°C, 20°C, 30°C, 40°C, 50°C and 60°C with the value of 2.5Ω load resistance. The measurements with different loads were done at 20°C, and 30Ω, 15Ω, 7.5Ω, 3.75Ω, 2.5Ω, 1.88Ω and 1.5Ω resistors were used. Each measurement was performed at least twice. Discharging a 850mAh battery with 2.5Ω load resistance takes approximately 1 hour and 40 minutes, with 1.5Ω it is one hour. Discharging with 30Ω took twenty hours. These measurements were time consuming and therefore they needed to be computer controlled.

During one measurement four batteries were measured of the same value of load resistance and had the same temperature conditions. The four batteries were packaged and immersed into a temperature controlled water basin. The required water temperature was controlled by the T3Ster thermal transient tester's THERMOSTAT device [6] with a variance of 0.01°C. The terminals of the battery block from the THERMOSTAT were connected to a PCB, which consisted of the preset load resistors, the switching devices and the control LEDs. This panel was connected to the measuring equipment. We used the T3Ster thermal transient tester [5] to measure the four batteries simultaneously, with the resolution of the measuring channels about 200mV/1024. The measured data were collected by the PC, which was directly connected to the tester. It also controlled the whole measurement processing and switched the loads off from a battery if it became fully discharged.

After evaluating the measured results with this software several observations were done. A typical discharge curve consists of four different parts: (1) the initial transient, (2) the hold-period, when the voltage has got only a very small decrease during a longer period of time, (3) the exhausting transient, when voltage begins to drop and suddenly hurtles down. It reaches the preset threshold voltage very quickly, where the measurement control circuit switches off the load from the battery's terminals. (4) At this moment, the output voltage jumps to a higher value and a slow self-recharging period begins with further voltage increase (see Fig. 1.).

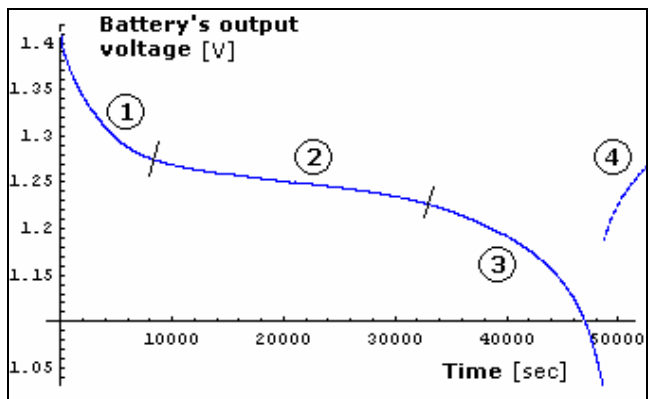

**Fig. 1. The four typical parts of the Ni-MH battery's discharge curve**

The measurement results revealed that varying load and temperature conditions shows significant dependence on the battery's parameters, as lifetime, operating voltage, curve's slope, and resultant capacity. Fig. 2 shows the discharge curves for four different temperature values.

The measured discharge curves were approximated with a proper mathematical function. We examined more possibilities to approach the curves, linear, exponential and hyperbolic functions were compared. The best possibility proved the hyperbolic approaching. For these discharging curves, the sum of one linear and two hyperbolic curves fits the best. The chosen approximation function is the following:

$$f(x) = \frac{A}{B+x} + \frac{C}{D+x} + E \cdot x + F$$

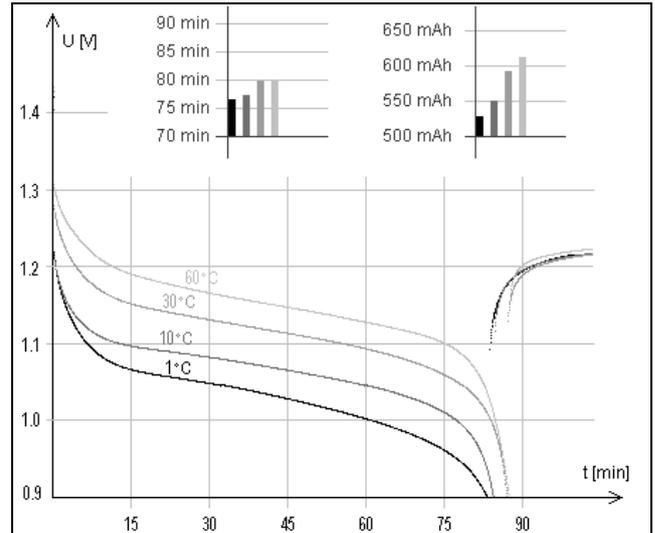

**Fig. 2. Ni-MH discharge curves at 1°C, 10°C, 30°C and 60°C, under constant resistance load**

The approaching function has got 6 parameters: A, B, C, D, E and F. The battery's properties appearing in the approximating function's parameters, so the parameters describe the battery and also the load and temperature dependence represented in the parameter's vary on these conditions.

The model creation process by following this principle now consists of the following steps: calculate all the fitting functions with the same formula (sum of 2 hyperbolic and 1 linear expression) for all of the measured discharge curves. Then we get the parameters of all the fitting functions in each condition cases (load and temperature). Having this information then the load and temperature dependent fitting functions have to be found for each parameter.

After calculating the fitting functions for the parameters, we got the following results: each parameter on temperature dependence show linear coherence, while for the load dependence, the linear parameters (E and F) shows linear coherence, and all the other shows hyperbolic.

Replacing the parameters with their calculated fitting curves, the model is complete.
The biggest disadvantage of this old model is that it calculates only with constant loads for the whole discharge period, and does not take the effects of pulsing current loads into consideration.

### 3. THE NEW MEASUREMENT SETUP

To overcome the limitations of this model we started to create a new measurement setup to be able to measure the batteries under not only constant resistance loads. As we planned a new design for the board, we also tried to remake all tasks automatic which were possible.

The measurement setup has been expanded with some new important parts. First of all, a microcontroller has been added. The discharging period is no longer controlled by the PC, but directly by this microcontroller. A new, more accurate voltage measuring part has been installed also on the board. A 4.096V voltage reference and a low-speed low-voltage 16-bit analog-to-digital converter IC takes also place on the new board, which is controlled by the microcontroller, and measures the battery's clamp voltage.

The other important change on the board is that the battery can also discharged under a preset constant current, and this current can also set to be time-pulsing, instead of constant resistor load. During the discharge process the output voltage of the measured battery is gradually decreasing, which means that if the battery is loaded with a constant resistance, current decreases also with voltage. To keep current stay in a proposed value we used a simple circuit with an operational amplifier, a transistor and an accurate resistor.

The pulsing load consists of the same cycles following each other. The magnitude of the load current can vary in every millisecond. The whole period consists of *n* impulses (including the idle time's zero impulses), which can be set from *1* to infinity, considering of course the capacity of the microcontroller's memory. With this program the waveform of the pulsing cycles can be determined, and the microcontroller is responsible for realizing this load program for the battery. The typical length of the cycles is in the range of 1-30 sec. Typically, each cycle has some initial current impulses, and then the remaining part of the period is a longer idle time, with no current drawn from the battery, as it can be seen in Fig. 3.

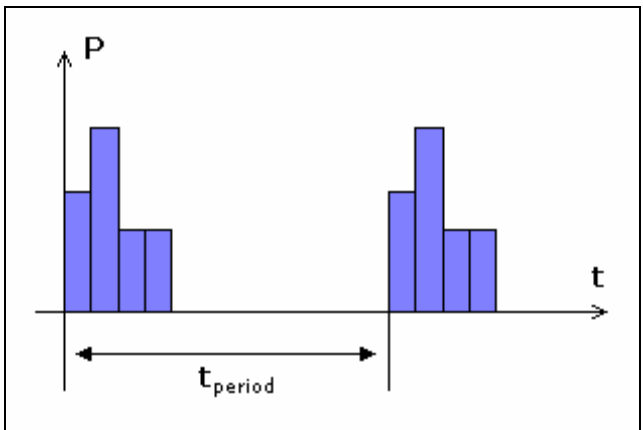

**Fig. 3. A typical pulsing load period consist of some current impulses and a longer idle time**

Considering that a mobile application rarely consumes more power than 50mW, our board has been designed to be able to generate 255 different impulses in the range of 1mW to 255mW. The 256th value is 0mW. This means that a single pulse period consists of impulses from this range.

The waveform of the load can be designed easily on the PC with a little software developed for this purpose, and then downloaded onto the measurement board's microcontroller. For the measurement, we use four AAA size Ni-MH rechargeable batteries, with the capacitance of 800mAh and 850mAh.

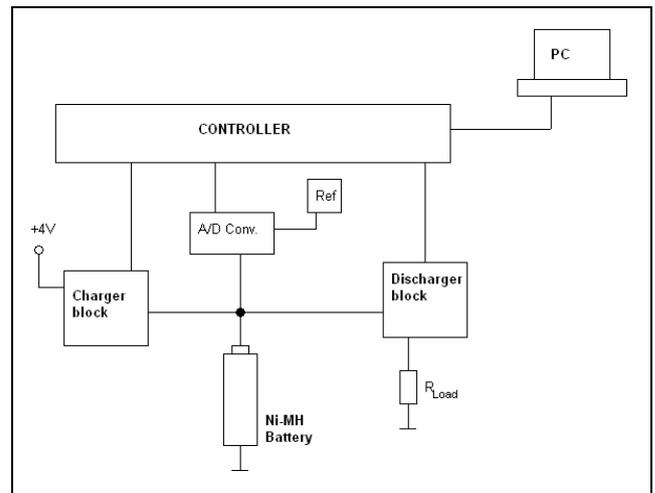

**Fig. 4. Block diagram of the measurement setup**

The block diagram of the new measurement setup is shown in Fig. 4. The measured data was collected by the PC. The PC had the task to collect and store the data for a later evaluation. The measurement board is connected to the PC through serial port, measured data is sent by the microcontroller on COM port. At the PC's side, there is a single data collecting and storing software, which has to be launched with the panel to catch the data. During the discharge period the measured output voltage is displayed for each channel and a diagram draws continuously the measured voltage-time curves. Results are compiled to a single text-format file.

The measurement process has to be preprogrammed before the measurement is launched. For the first step every measurement process has to be designed, and then compiled to be able to implement it in the microcontroller. After then the prepared program has to be uploaded to the microcontroller. The microcontroller manages the discharging process and also measures the voltage of the battery with the help of the A/D converter, and sends it towards the PC.

## 4. CHARGING BLOCK

The board has been expanded also with a charging circuit block. This part of the panel charges the plugged Ni-MH battery before the preprogrammed discharging cycle starts.

Charging the battery is realized as a simple constant current output circuit plugged onto the battery's clamp (similar to the discharging circuit), which pumps constant current into the battery. The control of the charging process is also the task of the microcontroller.

A fully discharged Ni-MH cell's output voltage is typically around 1.5V, when a 500mA current is flowing into it (under charging). During the charging cycle, the cell's voltage slowly increases, and when approaching the fully charged state voltage increments with greater gradient. When reaching the fully charged state, voltage is typically between 1.6-1.8V, and then the voltage increment stops and depending on the cell and the charging current, the voltage curve becomes horizontally flat or it can also show a small decrease (less than 2-3mV per minute). This is the point when the cell became fully charged and the charging current has to be switched off.

The whole charging cycle is realized in the microcontroller by software. Initially the charging current is switched on to charge the cell, and from this moment, the cell voltage is continuously polled. While the cell voltage is under 1600mV, we do not care about the gradient of the voltage curve, because some types of Ni-MH cells can also produce negative gradient voltage curves in the initial phase of the charging cycle unless they are still in fully discharged state.

The microprocessor's program polls the output voltage and compares every measured value to its previous one to calculate the curve's gradient.

The voltage drop in the curve at the end of the charging process (when battery has fully charged), is very small (only a few millivolts), and can hardly be detected, considering the noise which appears on the battery's output and also comes from the analog-digital conversion. To lower the noise, we use averaging, but it is not enough to completely avoid impulse-type noises, when for example noises come from the AC mains.

To create a secure algorithm for detecting the battery's voltage drop when it reaches the fully charged state, we used the sliding window methodology. After calculating whether the gradient of the slope is negative or not, this one-bit information is stored in an array, and before every new entry is stored (loaded) in the array, data is shifted one step and the last (oldest) value is dropped out. In our algorithm there are 16 places in the array. To save memory and realize shifting very easy, the 16 bit information is stored in one integer type variable, which can now act in software as a shift-register.

While voltage is increasing during the charging process, at each sampling moment the slope of the curve is positive, and zeroes are shifted into the register. When voltage starts constantly decreasing, zeroes changes to ones. Due to the noise, slope detection gives occasionally false values, which shifts also into the register. In our application, the voltage curve is determined negative gradient, when minimum 12 of the 16 bits are ones, and there are not more than only four zero bits. If this condition comes true, the cell is considered to be fully charged and charging process terminates.

After the charging cycle a 30 minutes long idle period is inserted every time, before starting the discharge cycle. This is also managed by the microcontroller.

## 5. MEASUREMENTS

To collect more knowledge about the battery's properties, several discharge measurements had to be done with the new above described measuring board. Each measurement was performed with pulsing, constant current loads. Measurements were done with different pulse-periods, load current values and duty cycles. Each measurement reproduced the Ni-MH battery's unique discharge curve shape.

Due to the continuously pulsing load the output voltage of the battery will not be continual any more, but it shows curves similar to the square wave. When the load is switched on, the output voltage drops, and after a while switching off the load, voltage jumps (back) to a higher value. This square wave superimposes at the original discharge time-voltage curve.

The voltage difference between the two states when load is switched on and off can be explained with the inner resistance of the battery. Evaluating the measurement results we diagnosed that the value of the inner resistance do not vary significantly, only in the short period when battery is getting fully discharged, an increase of inner resistance can be observed. A typical pulsing load discharge curve is presented in Fig. 5.

While the load is switched on, the battery's output voltage decreases. In pulsing load mode, the decrease comes from two effects. Obviously during the discharging the battery is loosing charge, and the output voltage decreases monotonic. This voltage decrease plays role only when the pulse width is longer. In pulsing mode, after each idle cycle period, when load is switched on again, the output voltage drop follows the event slower, and it draws another, hyperbolic monotonic voltage decrease.

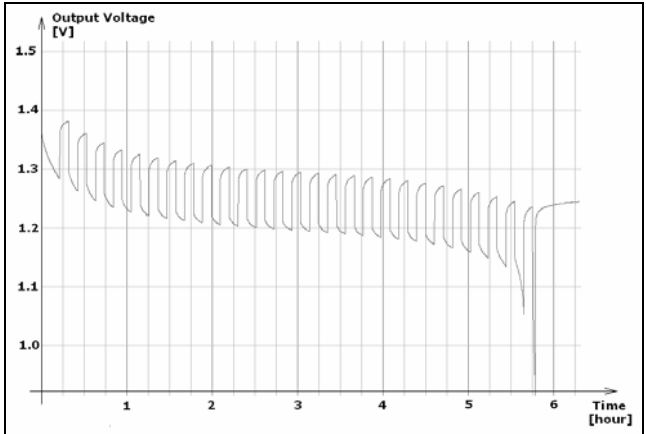

**Fig. 5. A typical pulsing load discharge curve. Pulse period=360s, duty=50%, load=250mA.**

As we draw charge out from the cell in each pulse cycle, at the end of every new cycle the output voltage is getting lower and lower, and also by the voltage increase during the idle period due to the recovery effect, the output voltage reaches lower and lower values.
Discharge process runs as long as the output cell voltage does not reach the lower threshold voltage 0.9V, when the load is switched off, and battery is considered to be fully discharged. Discharge measurements in practice are very long experiments; they can take from 6 hours to more than one day.

## 6. CREATING A MODEL

The measured results were evaluated with another computer application that was also developed at BME DED for this project. This program is capable to plot the selected measurement's time-voltage curves in a prescribed range and to zoom in them. The program can also calculate the effective measured capacity of the battery and display it in a chart. Time-derivative curves can also be plotted.

After evaluating the measured results with this software several observations were done. First of all, we diagnosed that each measurement reproduced the Ni-MH battery's unique discharge curve shape, but the length of the period, the determining voltages and other parameters shows significant differences.

Our ultimate goal is to create battery models, which will be able to provide the pulsing, constant current discharge transients for arbitrary load values, pulse period times and duty cycles. The models can be presented by mathematical approximating functions, and the parameters of these functions are dependent from the input parameters. From the several opportunities, we chose the hyperbolic approaching function. Typical Ni-MH discharge curves can be fitted with the sum of two hyperbolic and one linear expression, like:

$$f(x) = \frac{A}{B+x} + \frac{C}{D+x} + E \cdot x + F$$

This function has got six parameters. Two parameters (*A* and *B*) determine the first hyperbola, two determine the second (*C* and *D*), and the remaining two form a linear function. The first hyperbola's right side shapes the initial transient of the battery's discharge curve (the left side is dropped). The second hyperbola's left curve follows the curvature of the exhausting transient, the right part of this hyperbola is also ignored. The linear expression plays role between the two intervals, where the points are far enough from both asymptotes: in the hold-period. These six parameters of the final model (*A*, *B*, *C*, *D*, *E*, *and F*) will have the dependence of the pulsing time period, the duty cycle and the value of the load.

First we take the results (the voltage–time curves) of all completed measurements, and execute the fitting algorithm on them. For each measurement we get a unique approximating function, with its own values of the six parameters.

In the case of pulsing discharge loads, the measured curves can be fitted with two curves, an upper and a lower envelope curves. Upper curve approaches the voltage peaks at the end of the idle periods, while the lower fits to the lowest output voltage values at the end of the pulsing cycles (before load is switched off and idle cycle starts). To fit these curves, the above described hyperbolic formula is proper.

Both for the upper and lower envelope curves the A, B, C, D, E and F parameters have to be calculated, which also shows dependence on pulse width/period, duty cycle and load range. Therefore these parameters will not be constants, but they will be trivariate functions, and the three variables will be pulse period time, duty cycle and load. For example, the function of the A parameter looks like this:

$$A(L,p,d)$$

where *L* describes the value of load, *d* is the duty cycle, and *p* is the period in seconds.

After executing the fitting algorithm on the measured curves, and A, B, C, D, E and F parameters are calculated for each case, their dependence has to be examined.

Separate fitting functions have to be found for each parameters' several dependences. For example, knowing that the parameter E has a linear dependence on the load, the following formula is true:

$$E(L) = E_{La} \cdot L + E_{Lb}$$

where $E_{La}$ and $E_{Lb}$ are constant values. ($E_{La}$ is negative, while increasing the load, E will decrease.

## 7. CONCLUSION

Measurements with Ni-MH rechargeable batteries have been completed with different constant current pulsing condition to examine their discharge process.

The measurement was controlled by a board, which was developed by our team. Both charging and discharging blocks are implemented on the board. Measured data was collected by a PC. The results were evaluated with dedicated software, developed by the authors, and observed in the point of view of varying pulse period, duty cycle and the load value. The presented model predicts Ni-MH battery discharge transient curves under different pulsing current conditions. It is ready to be used in sensor network optimization applications.

## 8. ACKNOWLEDGEMENTS

This work was supported by the PATENT IST-2002-507255 Project of the EU and by the OTKA-TS049893 and the NKFP NAP 736-205/2005 BELAMI projects of the Hungarian Government.